\documentclass[aps,prl,reprint,groupedaddress]{revtex4-1}

\usepackage{graphicx}

\usepackage{xcolor}
\usepackage[normalem]{ulem}

\begin{document}


\title{Existence of Shapiro Steps in the Dissipative Regime in Superconducting Weak Links}


\author{Connor D.~Shelly}
\email[]{connor.shelly@npl.co.uk}
\affiliation{National Physical Laboratory, Hampton Road, Teddington, TW11 0LW, UK}

\author{Patrick See}
\affiliation{National Physical Laboratory, Hampton Road, Teddington, TW11 0LW, UK}

\author{Ivan Rungger}
\affiliation{National Physical Laboratory, Hampton Road, Teddington, TW11 0LW, UK}

\author{Jonathan M.~Williams}
\affiliation{National Physical Laboratory, Hampton Road, Teddington, TW11 0LW, UK}


\date{\today}

\begin{abstract}
We present measurements of microwave-induced Shapiro steps in a superconducting nanobridge weak link in the dissipative branch of a hysteretic current-voltage characteristic. We demonstrate that Shapiro steps can be used to infer a reduced critical current and associated local temperature. Our observation of Shapiro steps in the dissipative branch shows that a finite Josephson coupling exists in the dissipative state. Although the nanobridge is heated, our model shows that the temperature remains below the critical temperature.   This work provides evidence that Josephson behaviour can still exist in thermally-hysteretic weak link devices and will allow extension of the temperature range over which nanobridge based single flux quantum circuits, nanoSQUIDs and Josephson voltage standards can be used.
\end{abstract}


\maketitle



%

A superconducting weak link (WL) can be realised by creating a narrow constriction between two bulk superconducting electrodes. If the constriction dimensions are made sufficiently small (comparable to $3.5\xi$, where $\xi$ is the Ginzburg-Landau coherence length) then the WLs are expected to exhibit characteristic Josephson behaviour \cite{Likharev_weak_link}. Nanobridge constrictions can thus be used instead of traditional Josephson tunnel junctions based on oxide barriers, or superconductor-normal-superconductor (SNS) junctions. The majority of work in the area has focussed on development and optimization of micron-sized superconducting quantum interference devices (nanoSQUIDs), which are implemented using two WLs \cite{GranataReview2016}. NanoSQUIDs have application in single magnetic nanoparticle detection \cite{Hao_single_nanobead_2011}, scanning SQUID microscopy for imaging of nanoscale phenomena \cite{SSM_Nature_Nano_2013,SSM_RevSciInst_2012,Embon_SciRep_2015} and nano-electromechanical system (NEMS) readout \cite{Lolli_NEMS_2016,Patel_NEMS_nanoSQUID_2017}. Aside from magnetometer-based applications, WL Josephson junctions could be used in place of traditional junctions for single flux quantum (SFQ) circuits \cite{Likharev_Semenov_1991}, and Josephson voltage standards for metrology \cite{JMW_IEEE_2011,Jeanneret2009}. Weak links also have utility as Josephson elements in qubits and parametric amplifiers \cite{Tholen_2009,Levenson_APL_2011,Tholen_EPJ-QT_2014,Vijay_PRL_2009} as well as for single quasiparticle trapping and counting \cite{Levenson_QPTrap_2014}.

In general, hysteresis is observed in the current-voltage characteristics (IVC) of WLs. Unlike conventional tunnel junctions, where the hysteresis can be explained by capacitance in the resistively and capacitively shunted junction (RCSJ) model \cite{Tinkham}, the origin of hysteresis in WL junctions is attributed to heating, and subsequent thermal runaway of the junction \cite{Skocpol_hotspot}, similar to that observed in SNS junctions \cite{Peltonen_SNS_2008,Krasnov_PRB_2007}. This situation was first described by Skocpol, Beasley, and Tinkham (SBT) \cite{Skocpol_hotspot} who stated that as the bias current $I_\mathrm{dc}$ applied to the WL is increased above the critical current $I_\mathrm{c}$ a `hotspot' region in the WL forms, in which the local temperature exceeds the critical temperature $T_{\mathrm{c}}$. When $I_\mathrm{dc}$ is then reduced the hotspot is maintained by Joule heating. The WL is only able to return to the superconducting state when $I_\mathrm{dc}$ is reduced to below the retrapping current $I_\mathrm{r}$, where $I_\mathrm{r}<I_\mathrm{c}$. In recent years further refinements have been made to the SBT model by inclusion of a temperature dependent thermal conductivity at temperatures below $T_{\mathrm{c}}$ \cite{Hazra,Hazra_Kirtley_2015}, and extension of the model to millikelvin temperatures \cite{Blois_2017}. In addition, a significant amount of recent work has been carried out to understand and reduce the hysteresis in the IVC \cite{Kumar_PRL_2015,Kumar_SUST_2015,Arnaud_paper}.

Weak link thermal models \cite{Skocpol_hotspot, Hazra, Blois_2017} indicate that with a sufficiently large bias current the temperature in the WL can exceed $T_\mathrm{c}$, in some cases the $T>T_\mathrm{c}$ region is predicted to extend several micrometers into the electrodes. Indeed, Kumar \textit{et al}.~present a device-state diagram for WL-based nanoSQUIDs showing that at $T<T_{\mathrm{H}}$ (where $T_{\mathrm{H}}$ is the crossover temperature between the hysteretic and non-hysteretic regimes) and $I_{\mathrm{dc}}>I_{\mathrm{r}}$ the WLs and the micron-scale leads are in the resistive state \cite{Kumar_PRL_2015}. Preliminary nanoSQUID measurements in the hysteretic regime showed no magnetic flux dependence of the retrapping current \cite{Kumar_PRL_2015,Hazra_Kirtley_2015}. However, Biswas \textit{et al}.~have recently demonstrated that thermally-optimized nanoSQUIDs do exhibit magnetic flux dependence of $I_{\mathrm{r}}$ \cite{Biswas_PRB_2018}, indicating that the Josephson coupling does not completely vanish in the dissipative state. The Josephson effect can also be demonstrated through the observation of microwave-induced Shapiro steps \cite{SidneyShapiro}. We have previously observed Shapiro steps in WLs operated in the non-hysteretic regime \cite{Shelly_2017} and they have also been found in long nanowires when driven into the `phase slip center' regime \cite{Dinsmore_APL_2008,Bae_NJP_2012}.

In this Letter we demonstrate Josephson behaviour in hysteretic nanobridge WL junctions by observation of Shapiro steps, and combine the experimental data with our model to estimate the local temperature of the WL. To do this we measure the IVC of the WL at a temperature $T<T_{\mathrm{H}}$ whilst applying a radio-frequency (rf) current. Notably, we observe Shapiro steps on the dissipative branch of the hysteretic IVC previously thought to be in the fully normal state where it was assumed that the nanobridge and parts of the electrode have $T>T_{\mathrm{c}}$.

The WLs are fabricated by electron-beam lithography (EBL) and dry etching. A niobium film of thickness 150\,nm is sputtered onto a silicon substrate on top of which a 30\,nm thick aluminium film to be used as a hard mask is defined by EBL and thermally deposited by lift-off. An array of 10 nanobridges is defined to a width of 40\,nm and a length of 100\,nm. The niobium is then dry etched into the silicon substrate using a CHF$_{3}$/SF$_{6}$ plasma. The aluminium hard mask is left on.

 \begin{figure}
 \includegraphics[width=1\columnwidth]{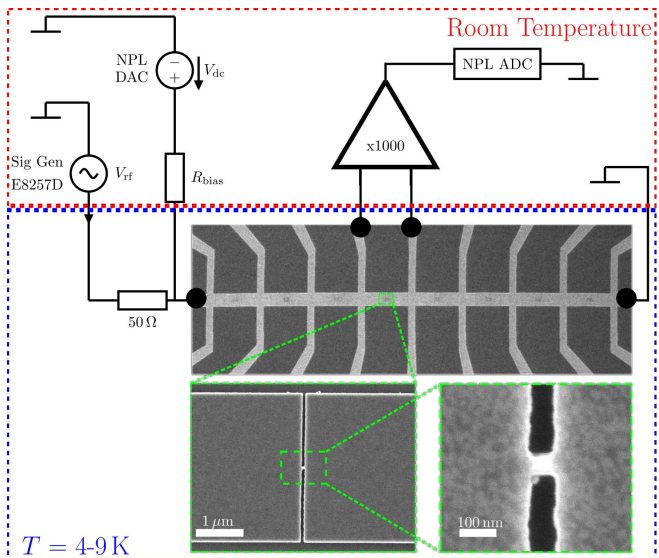}%
 \caption{Scanning electron micrographs of a niobium WL array and measurement schematic. To measure a single WL IVC a bias current is driven through the entire array via the NPL digital-to-analog converter (DAC) voltage source and a bias resistor. The voltage across an individual WL is measured using the NPL analog-to-digital converter (ADC). To investigate the influence of rf irradiation an rf current is applied to the entire array using an rf synthesizer and on-chip 50\,$\Omega$ resistor. \label{SEM_sketch}}
 \end{figure}

Electrical measurements of the WLs are carried out in a $^{4}$He dip Dewar. The temperature is varied between 4-9\,K by varying the position of the probe in the gas column. The IVC are measured in a four-terminal configuration using an optically isolated measurement unit optimised for high-precision electrical metrology \cite{Williams_Curve_Tracer_2009} designed at the National Physical Laboratory (NPL). To observe Shapiro steps, the WL is biased with an rf current. The scaling factor between the applied rf voltage from the synthesizer $V_{\mathrm{rf}}$ and the rf current that reaches the device $I_{\mathrm{rf}}$ is determined from two IVCs, see SM. A scanning electron micrograph and measurement schematic are shown in Figure \ref{SEM_sketch}.

\begin{figure}
 \includegraphics[width=1\columnwidth]{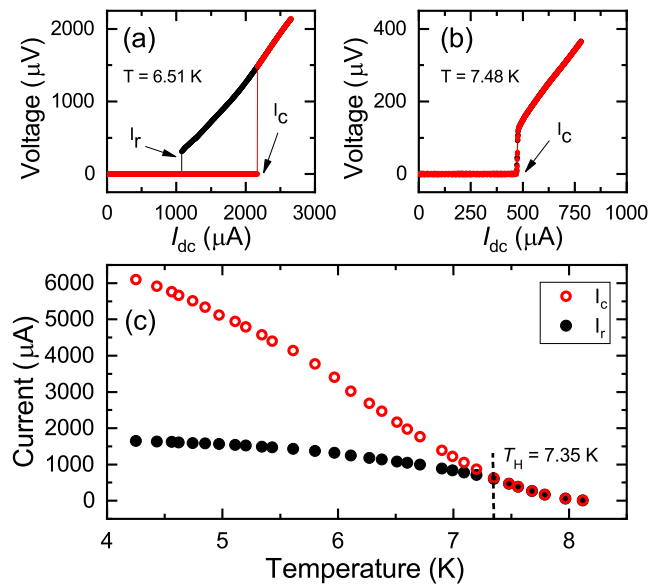}%
 \caption{(a) IVC of a weak link nanobridge measured at a bath temperature of 6.51\,K exhibiting thermal hysteresis. (b) IVC of weak link nanobridge at a bath temperature of 7.48\,K showing no hysteresis. (c) Critical current and retrapping current measured at different bath temperatures. Hysteresis occurs at $T<T_{\mathrm{H}}$. For this sample $T_{\mathrm{H}}\approx 7.35\,\mathrm{K}$.  \label{I_vs_T}}
 \end{figure}

Typical IVCs measured without rf current are shown in Figure \ref{I_vs_T} at different temperatures. The behaviour is qualitatively similar to that observed previously by the authors \cite{Shelly_2017} and elsewhere \cite{Hazra,Blois_2017,Arnaud_paper} showing that both $I_{\mathrm{c}}$ and $I_{\mathrm{r}}$ are temperature dependent and that the hysteresis disappears above $T_{\mathrm{H}}$.

\begin{figure}
  \includegraphics[width=1\columnwidth]{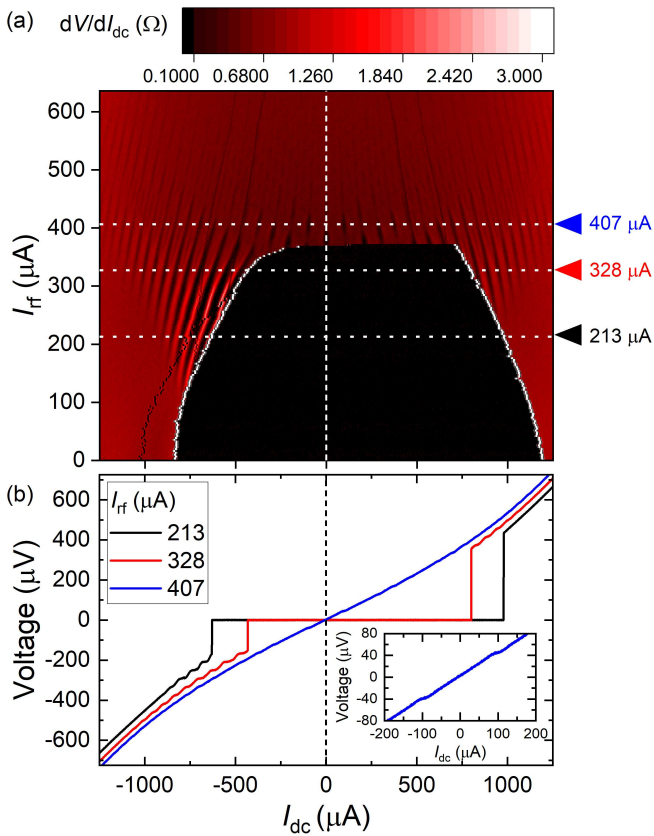}%
  \caption{(a) Differential resistance colourmap as a function of $I_{\mathrm{dc}}$ and $I_{\mathrm{rf}}$ measured at a bath temperature of 7\,K (hysteretic regime). The darker regions indicate flat features in the IVC corresponding to Shapiro steps at the expected voltages for $f_{\mathrm{rf}}=20\,\mathrm{GHz}$. (b) Selected IVC traces: At $I_{\mathrm{rf}}=213\,\mu\mathrm{A}$ steps are observed as $I_\mathrm{dc}$ is swept down from negative values to zero. $I_{\mathrm{rf}}=328\,\mu\mathrm{A}$ steps are seen on both branches (as current is swept from negative values to zero, and as current is swept from zero to positive values). When the applied rf is of sufficent amplitude the nanobridge no longer shows any evidence of hysteresis in the IVC as shown by the trace at $I_{\mathrm{rf}}=407\,\mu\mathrm{A}$. (inset) $I_{\mathrm{rf}}=407\,\mu\mathrm{A}$ trace over a smaller $I_{\mathrm{dc}}$ range showing the first observable step (n=1 step occurring at $V\approx41.4\,\mu\mathrm{V}$). \label{colourmap_635}}
  \end{figure}

Figure \ref{colourmap_635}(a) shows a differential resistance map obtained by numerically differentiating IVCs measured with increasing rf current (differential conductance map shown in SM). The differential resistance map shows the evolution of the IVC as a function of the applied rf current where the dark regions of the colourmap indicate a plateau in the IVC. These plateaus can be seen in the IVC traces and form at the expected Josephson voltage-frequency relation voltages $V=n(hf/2e)$, where $f=20\,\mathrm{GHz}$ is the frequency of the rf current. Figure \ref{colourmap_635}(b) shows three traces from the differential resistance map. The trace at $I_{\mathrm{rf}}=213\,\mu\mathrm{A}$ shows Shapiro steps on the down sweep (current from negative $I_\mathrm{dc}$ to zero) sweep of the dissipative branch of the hysteretic IVC. The trace at $I_{\mathrm{rf}}=328\,\mu\mathrm{A}$ shows that Shapiro steps appear on both the up sweep (current from zero to positive $I_\mathrm{dc}$) and the down  sweeps of the IVC. When sufficiently large rf currents are applied, as shown by the trace taken at $I_{\mathrm{rf}}=407\,\mu\mathrm{A}$, the hysteresis in the IVC disappears and the WL behaves as a non-hysteretic junction whilst still exhibiting Shapiro steps. This behaviour is qualitatively similar to that observed by de Cecco \textit{et al}.~in SNS Josephson junctions \cite{DeCecco2016}. 
The existence of Shapiro steps in the hysteretic IVC (at $I_{\mathrm{rf}}<372\,\mu\mathrm{A}$) on both the up and down sweeps indicates that the WL is not in a fully dissipative state but instead provides evidence of a finite Josephson supercurrent existing in the dissipative state in agreement with the recently observed retrapping current modulation \cite{Biswas_PRB_2018}. Similar to de Cecco \textit{et al}.~\cite{DeCecco2016}, we observe that at sufficiently high $I_{\mathrm{rf}}$ ($>372\,\mu\mathrm{A}$) the IVC become non-hysteretic, likely due to the temperature of the junction at $I_{\mathrm{dc}}=0$ increasing above the crossover temperature ($T>T_\mathrm{H}$).

\begin{figure}
  \includegraphics[width=1\columnwidth]{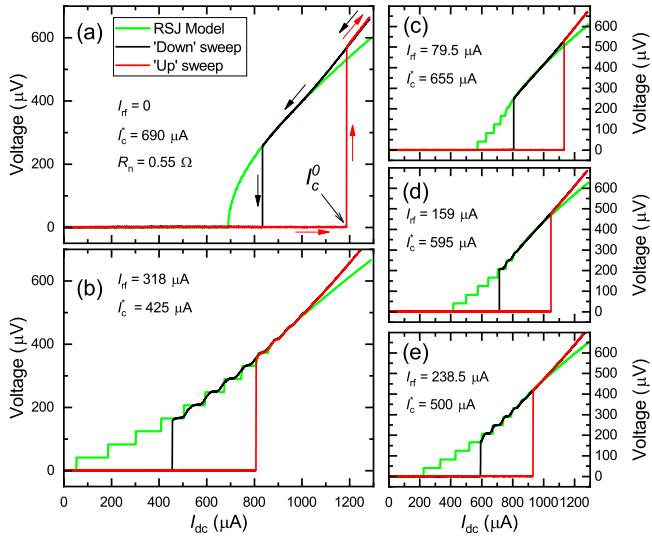}%
  \caption{IVC traces shown at different $V_{\mathrm{rf}}$ amplitudes. `Up' sweep shown in red, `Down' sweep shown in black, and RSJ model shown in green. (a) IVC at $I_{\mathrm{rf}}=0$ used to find $I_{\mathrm{c}}^{*}$ and $R_{\mathrm	{n}}$. The `up' sweep shows that as the initial state critical current $I_{\mathrm{c}}^{0}$ is reached the junction transitions to a lower critical current $I_{\mathrm{c}}^{*}$. During the `down' sweep the junction remains on this path until the Joule heating is no longer sufficient to stop the junction re-entering the fully superconducting regime. The RSJ model is used to determine $I_{\mathrm{c}}^{*}$ and  $R_{\mathrm{n}}$. $R_{\mathrm	{n}}$ is kept constant for the remainder of the analysis. (b) IVC with applied rf of $I_{\mathrm{rf}}=318\,\mu\mathrm{A}$. Our model uses $I_{\mathrm{c}}^{*}$ as the only fitting parameter and is able to reproduce the Shapiro step position and total number of steps for the full range of applied $I_{\mathrm{rf}}$ in the hysteretic region of the colourmap shown in Figure \ref{colourmap_635}. Measured IVC and RSJ model fits at (c) $I_{\mathrm{rf}}=79.5\,\mu\mathrm{A}$, (d) $I_{\mathrm{rf}}=159\,\mu\mathrm{A}$, and (e) $I_{\mathrm{rf}}=238.5\,\mu\mathrm{A}$. 
  \label{RSJ_traces}}
  \end{figure}

To explain our observation of Shapiro steps in the dissipative regime we first consider what happens to the critical current of a hysteretic WL. Figure \ref{RSJ_traces} shows IVCs at different rf currents. Both the up and down sweeps are shown. As $I_{\mathrm{dc}}$ is swept from zero to positive values the junction's initial state critical current $I_{\mathrm{c}}^{0}$ is reached, causing the junction to enter the dissipative regime. Due to Joule heating of the WL in this dissipative regime the local temperature increases to $T^{*}$. The reduced critical current associated with this temperature is thus described as $I_{\mathrm{c}}^{*}=I_{\mathrm{c}}(T^{*})$. The dissipative region of the IVC is now at this lower critical current. To determine this reduced  $I_{\mathrm{c}}^{*}$ we fit the dissipative region of the measured IVC by numerically solving the first-order differential equation describing the RSJ model with an applied rf current  \cite{Russer1972},
\begin{equation}
\frac{\hbar}{2eR_{\mathrm{n}}}\dot{\phi} + I_{\mathrm{c}}\sin(\phi) = I_{\mathrm{dc}} + I_{\mathrm{rf}}\sin(2\pi ft),
\label{eq:RSJ_rf}
\end{equation}
where we assume a sinusoidal current-phase relation (see SM).
We use the trace with $I_{\mathrm{rf}}=0$ in order to determine a value for the normal-state resistance $R_{\mathrm{n}}$ which we keep constant for all other IVC fits. The fit to the $I_{\mathrm{rf}}=0$ IVC is shown in Figure \ref{RSJ_traces}(a), with fitting parameters of $I_{\mathrm{c}}^{*}=690\,\mu\mathrm{A}$ and $R_{\mathrm{n}}=0.55\,\Omega$. Figures \ref{RSJ_traces}(b-e) show RSJ model fits to measured IVC at different $I_{\mathrm{rf}}$ values using $I_{\mathrm{c}}^{*}$ as the only fitting parameter. As $I_{\mathrm{rf}}$ is increased, the additional dissipated power leads to an increase in $T^{*}$ and corresponding reduction of $I_{\mathrm{c}}^{*}$. The RSJ model with $I_{\mathrm{c}}\rightarrow I_{\mathrm{c}}^{*}$  reproduces both the position of the Shapiro steps and the number of observable steps. At higher $I_{\mathrm{dc}}$ the data and model deviate, which we attribute to a further increase in temperature due to Joule heating and subsequent modification of $I_{\mathrm{c}}^{*}$ and $R_{\mathrm{n}}$. Our fitting also predicts the existence of steps below $I_{\mathrm{r}}$, which are inaccessible in our measurement. This is attributed to the reduction in Joule heating as $I_{\mathrm{dc}}$ is reduced, which leads to a decrease of the local temperature and thus an increase of $I_{\mathrm{c}}^{*}$ until it is in excess of $I_{\mathrm{dc}}$, at which point the WL enters the fully superconducting state again and the local temperature returns approximately to the bath temperature, $T_{\mathrm{bath}}$.

\begin{figure}
  \includegraphics[width=1\columnwidth]{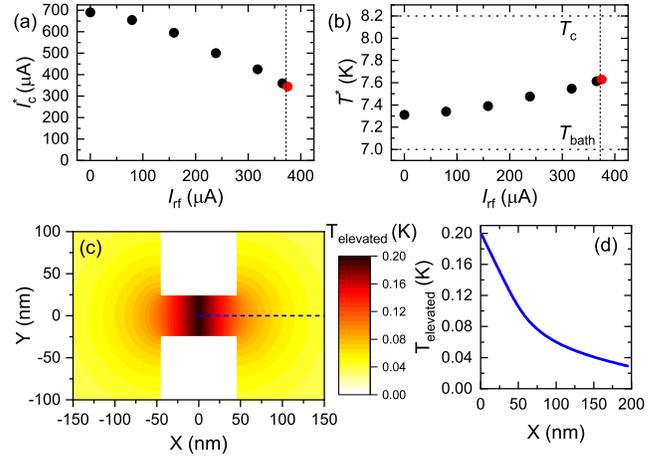}%
  \caption{(a) Best fit of $I_{\mathrm{c}}^{*}$ at different $I_{\mathrm{rf}}$. As $I_{\mathrm{rf}}$ is increased, $I_{\mathrm{c}}^{*}$ reduces. (b) Temperature inferred from value of $I_{\mathrm{c}}^{*}$ and the $I_{\mathrm{c}}(T)$ shown in Figure \ref{I_vs_T}(c). Dashed vertical line in both graphs refers to the $I_{\mathrm{rf}}$ value beyond which the IVC are non-hysteretic (i.e., $I_{\mathrm{rf}}>372\,\mu\mathrm{A}$ in Figure \ref{colourmap_635}). The fits to the RSJ model beyond this line (shown in red) are harder to achieve due to reduced step contrast. (c) Modelled temperature elevation in the nanobridge and electrodes. Power dissipation calculated from Shapiro step 6 of the IVC data in Figure \ref{RSJ_traces}(b). The temperature elevation at the centre of the nanobridge is approximately 0.2\,K. (d) Elevated temperature along the length of the nanobridge extending into the 150\,nm into the electrodes.= (data taken along blue dotted line in (c)).
  \label{T_Ic}}
  \end{figure}

The best fit values of $I_{\mathrm{c}}^{*}$ found at each $I_{\mathrm{rf}}$ are shown in Figure \ref{T_Ic}(a). As $I_{\mathrm{rf}}$ is increased, the value of $I_{\mathrm{c}}^{*}$ reduces as discussed above. The vertical dashed line in the figure denotes the crossover from hysteretic to non-hysteretic junction behaviour. After this point the steps in the IVC have less contrast and fitting is done using the numerically differentiated data. Only one point in this region is fitted to demonstrate that there is no large discontinuity beyond the crossover line. An estimate of the local temperature of the WL is made using the $I_{\mathrm{c}}(T)$ data from Figure \ref{I_vs_T}(c) and the results are shown in Figure \ref{T_Ic}(b). Note that our fitting procedure gives an estimate of $I_{\mathrm{c}}^{*}$, and thus the WL local temperature for values of $I_{\mathrm{dc}}$ in the vicinity of the hysteresis loop.

In a WL operated in the dissipative regime the temperature is not constant, but is expected to be highest in the center of the bridge, and to decrease within the electrodes. The estimated $T^{*}$ therefore is the local WL temperature and corresponds to the equivalent bath temperature of a WL with critical current $I_{\mathrm{c}}^{*}$ at $I_{\mathrm{rf}}=0$ (see Figure \ref{I_vs_T}(c)). We are interested in this local temperature at the centre of the bridge in order to determine whether the bridge is heated into a normal state (SNS) junction, or remains in a superconducting state (SS'S) where S' represents a superconducting region with modified properties. To determine the full temperature profile in the immediate proximity of the bridge we employ a finite element numerical method. The model used assumes a radial temperature distribution for distances further into the electrodes \cite{Skocpol_hotspot,Hazra}. The radial distribution is modelled with a modifed Bessel function of the second kind dependence and tends to $T_{\mathrm{bath}}$ at large distances (see SM for details of the model). Heat flow to the substrate is considered to be small in the bridge and its immediate proximity, but becomes significant deeper into the electrodes and is thus incorporated in the Bessel function dependence in the model. Figure \ref{T_Ic}(c) shows the results of the numerical modelling for the nanobridge at a bath temperature of 7\,K. The power dissipation input to the model is found from the experimental IVC data of the sixth Shapiro step in Figure \ref{RSJ_traces}(b), thus $P=150\,\mathrm{nW}$. The temperature elevation above $T_{\mathrm{bath}}$ from this modelling is found to be 0.2\,K (see Figure \ref{T_Ic}(c-d)). This temperature elevation is lower than the inferred $T^{*}$ shown in Figure \ref{T_Ic}(b) but this can be accounted for by including power dissipation in the on-chip $50\,\Omega$ resistor due to the rf current (see SM). The thermal conductivity inferred from the electrical measurements ($\kappa=5.2\,\mathrm{W\,m^{-1}\,K^{-1}}$) and the thermal conductivity found from measurements of heat flow in the bridge ($\kappa=4.3\,\mathrm{W\,m^{-1}\,K^{-1}}$) agree within $20\,\%$ (see SM).  Therefore, from both the inferred $T^{*}$ from our Shapiro step data, and the thermal modelling of our nanobridge we demonstrate that the dissipative branch of the IVC (in which Shapiro steps are observed) is an SS'S junction with a reduced critical current $I_{\mathrm{c}}^{*}$ and not an SNS junction. This result is contrary to some previous work in which a normal region was inferred to extend deep (many microns) into the electrodes, with no Josephson behaviour observed on the retrapping branch \cite{Arnaud_paper,Hazra_Kirtley_2015,Kumar_PRL_2015,Blois_2017}.

In conclusion we present experimental evidence of a finite Josephson supercurrent existing in the dissipative state of WL Josephson junctions demonstrated by the existence of Shapiro steps on the retrapping branch of the device IVC. We use the RSJ model in combination with the reduced critical current $I_{\mathrm{c}}^{*}$ to describe the evolution of the Shapiro steps over the full range of our hysteretic data, and to infer a local temperature of the WL. From the existence of Shapiro steps, a non-vanishing Josephson supercurrent, and our thermal modelling we show that our bridge temperature does not exceed $T_{\mathrm{c}}$.

Importantly, the existence of a Josephson supercurrent also demonstrates that WLs may be operated as Josephson junctions even in the dissipative state. This has relevance to the operation of hysteretic WL-based nanoSQUIDs, as well as demonstrating that rf irradiation can be used as a probe of Josephson behaviour in the dissipative regime of single WLs.  It is also critical to the understanding of WLs for use in applications such as SFQ circuits and Josephson voltage standards where response to high-frequency (GHz) pulses are important. The Shapiro steps can be used as a tool with which to investigate WL behaviour as well as informing the optimisation of WL junctions and SQUIDs. Different geometries, materials, and thermal shunts can be investigated using this rf irradiation technique.


%



\begin{acknowledgments}
This work is funded as part of a feasibility study funded by Innovate UK: Project number 102677. Co-funding supported by the UK NMS Electromagnetics and Time Programme and the EPSRC. We thank P.~J.~Meeson, J.~Ireland, A.~Ya.~Tzalenchuk, V.~T.~Petrashov, J.~Burnett, and T.~Lindstr\"om for useful technical discussion, and J.~C.~Gallop and S.~E.~de Graaf for critical reading of the manuscript. We also thank J.~P.~Griffiths and G.~A.~C.~Jones (Cavendish Laboratory, University of Cambridge) for the electron beam lithography.
\end{acknowledgments}

\bibliography{CDS_Shapiro_Heating}

\end{document}